\documentclass[12pt,reqno]{amsart}
\usepackage{indentfirst, amssymb,amsmath,amsthm,hyperref}
\usepackage{physics,breqn}
\usepackage[mathscr]{euscript}

\usepackage[caption=false]{subfig}
\usepackage{graphicx}
\usepackage{tabulary}
\usepackage{epstopdf}

\newtheorem{theo}{Theorem}

\newtheorem{rem}{Remark}

\newtheorem{cor}{Corollary}
\newtheorem{defi}{Definition}

\newcommand{\be}{\begin{equation}}
	\newcommand{\ee}{\end{equation}}
\newcommand{\beas}{\begin{eqnarray*}}
	\newcommand{\eeas}{\end{eqnarray*}}
\newcommand{\bea}{\begin{eqnarray}}
	\newcommand{\eea}{\end{eqnarray}}

\numberwithin{equation}{section}

\begin{document}
	
\setlength{\unitlength}{1mm}\baselineskip .45cm
%\large
\setcounter{page}{1}
\pagenumbering{arabic}	
\title[]{Pseudo generalized Ricci-recurrent spacetimes and modified gravity }
	
\author[  ]{Krishnendu De *$^{1}$ and Uday Chand De$^{\,2}$  }

\address
{$^2$ Department of Mathematics,\\
 Kabi Sukanta Mahavidyalaya,
The University of Burdwan.\\
Bhadreswar, P.O.-Angus, Hooghly,\\
Pin 712221, West Bengal, India.\\ ORCID iD: https://orcid.org/0000-0001-6520-4520}
\email{krishnendu.de@outlook.in, krishnendu.de67@gmail.com }
\address
{$^1$ Department of Pure Mathematics, University of Calcutta, West Bengal, India. ORCID iD: https://orcid.org/0000-0002-8990-4609}
\email {uc$_{-}$de@yahoo.com, ucde1950@gmail.com}
	
\begin{abstract}
In this paper we introduce and characterize a pseudo generalized Ricci-recurrent spacetimes and produce an example to verify the existence of such a spacetime. Then we demonstrate that a conformally flat generalized Ricci-recurrent spacetime with certain condition is a pseudo quasi-Einstein spacetime. Besides, it is proved that a pseudo generalized Ricci-recurrent generalized Robertson-Walker spacetime represents a  perfect fluid spacetime. Lastly, we study the impact of this spacetime under $f(\mathcal{R},T^2)$ and $f\left(\mathcal{R}^{\ast}\right)$ gravity scenario and deduce several energy conditions.
\end{abstract}
\footnotetext {PACS: 04.50.Kd, 98.80.jk, 98.80.cq.}:	
%\subjclass[2020]{53C50, 83C10, 83C56, 83D05, 53Z05.}
\keywords{Pseudo generalized Ricci-recurrent spacetime; generalized Robertson-Walker spacetime; perfect fluid spacetime; modified gravity.\\
*Corresponding author.}

\maketitle

\section{Introduction}

Let $M^{n}$ ($n\ge4$) be a semi (or, pseudo)-Riemannian manifold and $g$ be its semi-Riemannian metric with signature $(p,m)$, such that $p+m=n$. $M^{n}$ endowed with $g$ is referred to be a Lorentzian manifold \cite{Neil} if $g$ is a Lorentzian metric with signature $(n-1, 1)$ or, $(1, n-1)$. Time-oriented Lorentzian manifolds are Lorentzian manifolds that accept a globally time-like vector field, also physically referred to as spacetime. A generalized Robertson-Walker (shortly, GRW) spacetime is a Lorentzian manifold of dimension $n$ $\left(n\geq4\right)$ which can be shaped by a warped product $-I\times_{\varPsi^{2}}\stackrel{\ast}{M}$, $\stackrel{\ast}{M}$ is an $\left(n-1\right)$-dimensional Riemannian manifold, $I \in \mathbb{R}$ (set of real numbers) is an open interval, and $\varPsi>0$ is the warping function. This notion was presented by Al\'ias et al. \cite{alias1} in $1995$. In particular, GRW spacetime becomes Robertson-Walker ( shortly, RW) spacetime if we assume that $\stackrel{\ast}{M}$ is a 3- dimensional Riemannian manifold and is of constant sectional curvature. For more details about GRW spacetimes, we refer (\cite{alias1}-\cite{S98}).\par

A $4$-dimensional Lorentzian manifold $\mathrm{M}^{4}$ is described as a perfect fluid spacetime (shortly, PFS) if the Ricci tensor $\mathcal{R}_{lk}$ fulfills
\begin{equation}\label{1.1}
\mathcal{R}_{lk}=\alpha g_{lk}+\beta u_{l}u_{k},
\end{equation}
$\alpha$, $\beta$ being scalars (not simultaneously zero) and the velocity vector $u_{k}$ denotes the unit time-like vector \{that is, $u_{k}u^{k}=-1$, $u^{k}=g^{lk}u_{l}$\}. The matter field in general relativity (in short, GR) is described by $T_{lk}$, which is called the energy momentum tensor (in short, EMT). Since heat conduction term and stress terms corresponding to viscocity does not occur, the fluid is referred to as perfect \cite{HE73}. The EMT \cite{Neil} for a PFS has the following form
\begin{equation}\label{1.2}
T_{lk}=\left(\sigma+p\right)u_{l}u_{k}+p g_{lk},
\end{equation}
$p$, $\sigma$ being the isotropic pressure and the energy density, respectively. If $\sigma=p$, then the PFS is named stiff matter fluid. If $\sigma+p=0$, $p=0$ and $\sigma=3p$, then the PFS is called as the dark energy epoch of the Universe, dust matter fluid and radiation era \cite{chav}, respectively. Without a cosmological constant, the Einstein's field equations (shortly, EFEs) are stated as
\begin{equation}\label{1.3}
\kappa T_{lk}-\mathcal{R}_{lk}+\dfrac{\mathcal{R}}{2}\,g_{lk}=0
\end{equation}
in which $\kappa$ and $\mathcal{R}$ indicate the gravitational constant and Ricci scalar, respectively. %The above expression \eqref{1.1} of $\mathcal{R}_{lk}$ is obtained from EFE utilizing \eqref{1.2}.
 Equations \eqref{1.1}, \eqref{1.2} and \eqref{1.3} infer that
\begin{equation}\label{1.4}
	\alpha=\kappa\left(\dfrac{\sigma-p}{2}\right)\quad\mathrm{and}\quad\beta=\kappa\left(\sigma+p\right).
\end{equation}

In $\mathrm{M}^{4}$, the Weyl conformal curvature tensor $\mathcal{C}_{lijk}$ is demonstrated by
\begin{equation}\label{1.5}
	\mathcal{C}_{lijk}=\mathcal{R}_{lijk}-\dfrac{1}{2}\left\{g_{ij}\mathcal{R}_{lk}-g_{ik}\mathcal{R}_{lj}
+g_{lk}\mathcal{R}_{ij}-g_{lj}\mathcal{R}_{ik}\right\}
+\dfrac{\mathcal{R}}{6}\left\{g_{lk}g_{ij}-g_{lj}g_{ik}\right\}
\end{equation}
in which $\mathcal{R}_{lijk}$ stands for the curvature tensor.\par
It is well-circulated that \cite{E49}
\begin{equation}\label{1.6}
	\nabla_{l}\mathcal{C}^{l}_{ijk}=\dfrac{1}{2}\left\{\left(\nabla_{k}\mathcal{R}_{ij}-\nabla_{j}\mathcal{R}_{ik}\right)
-\dfrac{1}{6}\left(g_{ij}\nabla_{k}\mathcal{R}-g_{ik}\nabla_{j}\mathcal{R}\right)\right\}.
\end{equation}
%In \cite{H21}, the author studied a mixed super quasi-Einstein manifold and obtained
If the covariant derivative of $\mathcal{R}_{lk}$ is in the following form:
\begin{equation}\label{1.7}
\nabla_{i}\mathcal{R}_{lk}=u_{i}\mathcal{R}_{lk}+v_{i}g_{lk}+w_{i}\mathcal{D}_{lk},
\end{equation}
where $u_{i}$, $v_{i}$ and $w_{i}$ are non-zero covariant vectors, $\mathcal{D}$ is the $\left(0,2\right)$-type structure tensor of the manifold obeying
\begin{equation}\label{1.8}
\mathcal{D}_{lk}=\mathcal{D}_{kl},\quad g^{lk}\mathcal{D}_{lk}=0\quad\mathrm{and}\quad\mathcal{D}_{lk}u^{k}=0,
\end{equation}
then we named such an $n$-dimensional manifold as pseudo generalized Ricci-recurrent manifold, indicated by $P\left(GR_{n}\right)$. If $w_{i}=0$, then $P\left(GR_{n}\right)$ reduces to a generalized Ricci-recurrent manifold $\left(GR_{n}\right)$ \cite{DGK95}. For $v_{i}=w_{i}=0$, $P\left(GR_{4}\right)$ becomes a Ricci-recurrent manifold $\left(R_{n}\right)$ \cite{P52}. If $\mathcal{R}_{lk}$ obeys \eqref{1.7} and $u_{i}$ is a unit time-like vector, then $\mathrm{M}^{4}$ is called a $P\left(GR_{4}\right)$ spacetime. \par
Multiplying \eqref{1.7} with $g^{lk}$ and using \eqref{1.8}, we reach
\begin{equation}\label{1.9}
\nabla_{i}\mathcal{R}=\mathcal{R}u_{i}+4v_{i}.
\end{equation}

EFEs are not adequate to determine the late-time inflation of the cosmos without supposing the existence of certain unseen components that could account for the dark energy and dark matter origins. It is the primary source of inspiration for the extension to find higher order field equations of gravity.\par

Through the introduction of certain couplings between the geometrical quantities and the matter sector, the $f(\mathcal{R})$ theory of gravity has been further generalized. The non-minimally coupling between the matter lagrangian density and the curvature invariant has been proved in \cite{har}, which is known as the $f(\mathcal{R},L_m)$ theory of gravity. The corresponding Lagrangian can be modified by incorporating an analytic function of $T_{jk}T^{jk}$ in this generalisation procedure for the $f(\mathcal{R},L_m)$ theory. $f(\mathcal{R},T^2)$ gravity or energy-momentum squared gravity is the result of selecting the corresponding Lagrangian. In 2014, Katirci and Kavuk \cite{kat} originally put this theory, which permits the existence of a term in the action functional that is proportionate to $T_{jk}T^{jk}$. Additional research on this theory has been conducted by various researchers. Recent research (\cite{liu}, \cite{ak}, \cite{na}, \cite{ak1}) has shown that this modified theory has a number of cosmological applications.\par
Numerous alterations to EFEs have been constructed and extensively examined in (\cite{car}, \cite{kde1}). In (\cite{car1}, \cite{ea}), the authors show that Friedmann cosmological solutions have late-time accelerating attractors. They also demonstrate a specific model where the reconcile to GR theory is a polynomial function of $\mathcal{R}^2$, $\mathcal{R}_{jk}\mathcal{R}^{jk}$, and $\mathcal{R}_{ijkl}\mathcal{R}^{ijkl}$ quadratic curvature invariants. Here we investigate, another modification of $f(\mathcal{R})$ theory which was named $f\left(\mathcal{R}^{\ast}\right)$-gravity where $\mathcal{R}^{\ast}=\mathcal{R}_{jk}\mathcal{R}^{jk}$, introduced by Li et al. \cite{li}. The literature mentioned above makes it very evident that more attention needs to be paid to modified gravity, and there are still a lot of unanswered questions. Inspired by the foregoing investigations, this article is focused to investigate $P\left(GR_{4}\right)$ GRW spacetime satisfying $f(\mathcal{R},T^2)$ and $f\left(\mathcal{R}^{\ast}\right)$ gravity. In this paper, we choose a model $f\left(\mathcal{R}^{\ast}\right)=\ln(\mathcal{R}^{\ast})-\exp(-\mathcal{R}^{\ast})$, which is constructed to explain different Energy conditions (ECs). \par

This article is structured as:
In Section 2, an example of a $P\left(GR_{4}\right)$ spacetime is illustrated. In the next two Sections, we study a conformally flat $P\left(GR_{4}\right)$ spacetime and a $P\left(GR_{4}\right)$ GRW spacetimes. Finally, a $P\left(GR_{4}\right)$ GRW spacetime in $f(\mathcal{R},T^2)$ and $f\left(\mathcal{R}^{\ast}\right)$ gravity theory is considered.

\section{Example of a $P\left(GR_{4}\right)$ spacetime}
\noindent
Choose a Lorentzian metric $g$ on $\mathbb{R}^{4}$ described by %\cite{MD19}
\begin{equation}\label{2.1}
ds^{2}=g_{ij}dy^{i}dy^{j}=\left(dy^{1}\right)^{2}+\left(y^{1}\right)^{2}\left(dy^{2}\right)^{2}
+\left(y^{2}\right)^{2}\left(dy^{3}\right)^{2}-\left(dy^{4}\right)^{2},
\end{equation}
where $i,j=1,2,3,4$. Using \eqref{2.1}, we observe that the non-vanishing components of the metric tensor are
\begin{equation}\label{2.2}
g_{11}=1,\quad g_{22}=\left(y^{1}\right)^{2},\quad g_{33}=\left(y^{2}\right)^{2},\quad g_{44}=-1
\end{equation}
and the associated contravariant components are
\begin{equation}\label{2.3}
g^{11}=1,\quad g^{22}=\dfrac{1}{\left(y^{1}\right)^{2}}\,,\quad g^{33}=\dfrac{1}{\left(y^{2}\right)^{2}}\,,\quad g^{44}=-1.
\end{equation}
Utilizing equations \eqref{2.2} and \eqref{2.3}, here we determine the components (non-vanishing) of the Christoffel symbols, the curvature tensor and the Ricci tensor and they are
\begin{equation}\label{2.4}
\Gamma_{22}^{1}=-y^{1},\quad\Gamma_{33}^{2}=-\dfrac{y^{2}}{\left(y^{1}\right)^{2}}\,,\quad\Gamma_{12}^{2}
=\dfrac{1}{y^{1}}\,,\quad\Gamma_{23}^{3}=\dfrac{1}{y^{2}}\,,
\end{equation}
\begin{equation}\label{2.5}
\mathcal{R}_{1332}=-\dfrac{y^{2}}{y^{1}}\,,\quad\mathcal{R}_{12}=-\dfrac{1}{y^{1}y^{2}}
\end{equation}
and the symmetric properties lead to the other components.\par
The covariant derivative of non-vanishing Ricci tensor is written as
\begin{equation}\label{2.6}
\mathcal{R}_{12,1}=\dfrac{1}{y^{2}\left(y^{1}\right)^{2}}\quad\mathrm{and}\quad\mathcal{R}_{12,2}
=\dfrac{1}{y^{1}\left(y^{2}\right)^{2}}\,.
\end{equation}
The one-forms we select are as follows:
\begin{equation}\label{2.7}
u_{i}\left( y\right)  =
\begin{cases}
1, & $ when $ i = 4\\
0, &  $ otherwise, $
\end{cases}
\end{equation}
\begin{equation}\label{2.8}
v_{i}\left( y\right)  =
\begin{cases}
y^{1}, & $ when $ i=2\\
y^{2}, &  $ when $ i=3 \\
\,0, &  $ otherwise $
\end{cases}
\end{equation}
and
\begin{equation}\label{2.9}
w_{i}\left( y\right)  =
\begin{cases}
y^{2}, & $ when $ i=1\\
y^{1}, &  $ when $ i=2 \\
\,0, &  $ otherwise $
\end{cases}
\end{equation}
for all $y\in\mathbb{R}^{4}$.\par

We consider $\mathcal{D}_{ij}$ as follows:
\begin{equation}\label{2.10}
\mathcal{D}_{ij}\left( y\right)  =
\begin{cases}
\dfrac{1}{\left(y^{1}\right)^{2}\left(y^{2}\right)^{2}}\,, & $ when $ i = 1,\, j=2\\
\quad\quad0, &  $ otherwise $
\end{cases}
\end{equation}
for all $y\in\mathbb{R}^{4}$. It is enough to examine at the subsequent equations in order to confirm the relation \eqref{1.7}:
\begin{equation}\label{2.11}
\mathcal{R}_{12,1}=u_{1}\mathcal{R}_{12}+v_{1}g_{12}+w_{1}\mathcal{D}_{12}
\end{equation}
and
\begin{equation}\label{2.12}
\mathcal{R}_{12,2}=u_{2}\mathcal{R}_{12}+v_{2}g_{12}+w_{2}\mathcal{D}_{12}.
\end{equation}
The other cases holds trivially.
\begin{align*}
\text{Now, right hand side of \eqref{2.11}} &=u_{1}\mathcal{R}_{12}+v_{1}g_{12}+w_{1}\mathcal{D}_{12}\\
&=0\cdot\left(\dfrac{-1}{y^{1}y^{2}}\right)+0+y^{2}\cdot\dfrac{1}{\left(y^{1}\right)^{2}\left(y^{2}\right)^{2}}\\
&=\dfrac{1}{y^{2}\left(y^{1}\right)^{2}}=\mathcal{R}_{12,1}.
%	&= \text{L.H.S. of (44)}
\end{align*}
By applying the same deduction, it is possible to demonstrate that equation \eqref{2.12} is likewise true.\par
By virtue of \eqref{2.3} and \eqref{2.7} we find
\begin{equation}\label{2.13}
g^{ij}u_{i}u_{j}=-1,\quad\mathrm{and}\quad u^{i}\left(y\right)=g^{ij}u_{j}\left(y\right)=
\begin{cases}
	-1, & $ when $ i = 4\\
	\;\;0, &  $ otherwise $
\end{cases}
\end{equation}
Equations \eqref{2.10} and \eqref{2.13} together yield
\begin{equation}\label{2.14}
\mathcal{D}_{ij}u^{i}=0.
\end{equation}
Clearly, the trace$\left(\mathcal{D}_{ij}\right)=0.$ Therefore, $\left(\mathbb{R}^{4},g\right)$ is a $P\left(GR_{4}\right)$ spacetime.

\section{Conformally flat $P\left(GR_{4}\right)$ spacetimes}
\begin{defi} %\cite{S09}
A $\mathrm{M}^{4}$ is called a pseudo quasi-Einstein (in short, PQE) spacetime if $\mathcal{R}_{ij}$ satisfies the following:
\begin{equation}\label{3.1}
\mathcal{R}_{ij}=\alpha_{1}g_{ij}+\alpha_{2}u_{i}u_{j}+\alpha_{3}\mathcal{D}_{ij}
\end{equation}
in which $\alpha_{1},$ $\alpha_{2}$ and $\alpha_{3}$ are non-zero smooth functions.
\end{defi}
\noindent
For a conformally flat spacetime (that is, $\mathcal{C}_{hijk}=0$), it follows easily that \cite{E49} $\nabla_{l}\mathcal{C}^{l}_{ijk}=0$, that is,
\begin{equation}\label{3.2}
\nabla_{k}\mathcal{R}_{ij}-\nabla_{j}\mathcal{R}_{ik}=\dfrac{1}{6}\left\{g_{ij}\left(\nabla_{k}\mathcal{R}\right)
-g_{ik}\left(\nabla_{j}\mathcal{R}\right)\right\}.
\end{equation}
Equations \eqref{1.7}, \eqref{1.9} and \eqref{3.2} together imply
\begin{align}\label{3.3}
u_{k}\mathcal{R}_{ij}&+v_{k}g_{ij}+w_{k}\mathcal{D}_{ij}-u_{j}\mathcal{R}_{ik}-v_{j}g_{ik}
-w_{j}\mathcal{D}_{ik}\nonumber\\
&=\dfrac{1}{6}\left\{\mathcal{R}\left(u_{k}g_{ij}-u_{j}g_{ik}\right)+4\left(v_{k}g_{ij}-v_{j}g_{ik}\right)\right\}.
\end{align}
Multiplying \eqref{3.3} with $u^{k}$ and using \eqref{1.8}, we notice that
\begin{equation}\label{3.4}
\mathcal{R}_{ij}=\left\{\dfrac{\left(v_{k}u^{k}\right)}{3}+\dfrac{\mathcal{R}}{6}\right\}g_{ij}
+\dfrac{\mathcal{R}}{6}\,u_{j}u_{i}-\dfrac{1}{3}\,v_{j}u_{i}-u_{j}\mathcal{R}_{ik}u^{k}
+\left(w_{k}u^{k}\right)\mathcal{D}_{ij}.
\end{equation}
Suppose that corresponding to the eigenvector $u_{i}$, the eigenvalue of the Ricci tensor is $\theta$, that is, $\mathcal{R}_{ik}u^{k}=\theta u_{i}$, $\theta$ being a scalar, then the foregoing equation \eqref{3.4} becomes
\begin{equation}\label{3.5}
\mathcal{R}_{ij}=\left(\dfrac{f_{1}}{3}+\dfrac{\mathcal{R}}{6}\right)g_{ij}+\left(\dfrac{\mathcal{R}}{6}-\theta\right)u_{i}u_{j}-\dfrac{1}{3}\,v_{j}u_{i}+f_{2}\mathcal{D}_{ij},
\end{equation}
where $f_{1}=v_{k}u^{k}$ and $f_{2}=w_{k}u^{k}$.\\
Interchanging $i$ and $j$ in \eqref{3.5}, we reach
\begin{equation}\label{3.6}
\mathcal{R}_{ji}=\left(\dfrac{f_{1}}{3}+\dfrac{\mathcal{R}}{6}\right)g_{ji}
+\left(\dfrac{\mathcal{R}}{6}-\theta\right)u_{j}u_{i}-\dfrac{1}{3}\,v_{i}u_{j}+f_{2}\mathcal{D}_{ji}.
\end{equation}
Subtracting \eqref{3.6} from \eqref{3.5}, we obtain
\begin{equation}\label{3.7}
v_{i}u_{j}=v_{j}u_{i}.
\end{equation}
Multiplying \eqref{3.7} with $u^{j}$, we have
\begin{equation}\label{3.8}
v_{i}=-\left(v_{j}u^{j}\right)u_{i},
\end{equation}
that is,
\begin{equation}\label{3.9}
v_{i}=-f_{1}u_{i}.
\end{equation}
Equations \eqref{3.5} and \eqref{3.9} give us
\begin{equation}\label{3.10}
\mathcal{R}_{ij}=\left(\dfrac{f_{1}}{3}+\dfrac{\mathcal{R}}{6}\right)g_{ij}
+\left(\dfrac{f_{1}}{3}+\dfrac{\mathcal{R}}{6}-\theta\right)u_{i}u_{j}+f_{2}\mathcal{D}_{ij}.
\end{equation}
Hence, we write:
\begin{theo}
A conformally flat $P\left(GR_{4}\right)$ spacetime with $\mathcal{R}_{jk}u^{k}=\theta u_{j}$ is a PQE spacetime.
%pseudo quasi-Einstein spacetime.
\end{theo}

\section{$P\left(GR_{4}\right)$ GRW spacetimes}
\noindent
{\bf Theorem A.} \cite{survey} A $\mathrm{M}^{4}$ is a GRW spacetime iff it permits a unit torse-forming time-like vector $u_{i}$:
\begin{equation}\label{4.1}
\nabla_{k}u_{i}=\varphi\left\{g_{ki}+u_{k}u_{i}\right\}
\end{equation}
and $u_{i}$ is an eigenvector of $\mathcal{R}_{ij}$, that is,
\begin{equation}\label{4.2}
\mathcal{R}_{ij}u^{j}=\theta u_{j}
\end{equation}
in which $\varphi$ and $\theta$ are non-zero scalars.\par
Multiplying \eqref{1.7} with $u^{i}$ and using \eqref{1.8} infers
\begin{equation}\label{4.3}
\left(\nabla_{k}\mathcal{R}_{ij}\right)u^{i}=u_{k}\mathcal{R}_{ij}u^{i}+v_{k}u_{j}.
\end{equation}
Since $\nabla_{k}\left(\mathcal{R}_{ij}u^{i}\right)=\left(\nabla_{k}\mathcal{R}_{ij}\right)u^{i}
+\mathcal{R}_{ij}\left(\nabla_{k}u^{i}\right)$, \eqref{4.3} becomes
\begin{equation}\label{4.4}
\nabla_{k}\left(\mathcal{R}_{ij}u^{i}\right)-\mathcal{R}_{ij}\left(\nabla_{k}u^{i}\right)
=u_{k}\mathcal{R}_{ij}u^{i}+v_{k}u_{j}.
\end{equation}
Equations \eqref{4.1}, \eqref{4.2} and \eqref{4.4} together imply
\begin{equation}\label{4.5}
\varphi\mathcal{R}_{jk}=\theta\varphi g_{jk}+\theta_{k}u_{j}-\theta u_{k}u_{j}-v_{k}u_{j}.
\end{equation}
Interchanging $j$ and $k$ in \eqref{4.5}, we obtain
\begin{equation}\label{4.6}
\varphi\mathcal{R}_{kj}=\theta\varphi g_{kj}+\theta_{j}u_{k}-\theta u_{j}u_{k}-v_{j}u_{k}.
\end{equation}
Subtracting \eqref{4.6} from \eqref{4.5}, we have
\begin{equation}\label{4.7}
\theta_{k}u_{j}=\theta_{j}u_{k}+v_{k}u_{j}-v_{j}u_{k}.
\end{equation}
Multiplying \eqref{4.7} with $u^{j}$, we acquire
\begin{equation}\label{4.8}
\theta_{k}=\left(v_{j}u^{j}\right)u_{k}-\left(\theta_{j}u^{j}\right)u_{k}+v_{k},
\end{equation}
that is,
\begin{equation}\label{4.9}
\theta_{k}=\left(f_{1}-f_{3}\right)u_{k}+v_{k},\quad\mathrm{where}\quad f_{1}=v_{j}u^{j}\quad\mathrm{and}\quad f_{3}=\theta_{j}u^{j}.
\end{equation}
From \eqref{4.5} and \eqref{4.9}, it follows that
\begin{equation}\label{4.10}
\mathcal{R}_{jk}=\theta g_{jk}+(\dfrac{f_{1}-f_{3}-\theta}{\varphi})u_{j}u_{k}.
\end{equation}
Hence, we reach:
\begin{theo}
A $P\left(GR_{4}\right)$ $\mathrm{GRW}$ spacetime represents a $\mathrm{PFS}$.
\end{theo}
\noindent
In light of equations \eqref{1.1}, \eqref{1.4} and \eqref{4.10}, we acquire
\begin{equation}\label{4.11}
\kappa\left(\dfrac{\sigma-p}{2}\right)=\theta
\end{equation}
and
\begin{equation}\label{4.12}
\kappa\left(\sigma+p\right)=\dfrac{f_{1}-f_{3}-\theta}{\varphi}\,.
\end{equation}
Equations \eqref{4.11} and \eqref{4.12} together give
\begin{equation}\label{4.13}
\dfrac{p}{\sigma}=\dfrac{f_{1}-f_{3}-\theta-2\theta\varphi}{f_{1}-f_{3}-\theta+2\theta\varphi}\,.
\end{equation}
We observe that \eqref{4.13} implies $p=0$ for $f_{1}=f_{3}+\theta\left(1+2\varphi\right)$, $\sigma=3p$ for $f_{1}=f_{3}+\theta\left(1+4\varphi\right)$ and $\sigma+p=0$ for $f_{1}=f_{3}+\theta$, respectively. Hence, we obtain:
\begin{cor}
A $ P\left(GR_{4}\right)$ $\mathrm{GRW}$ spacetime represents a
\begin{enumerate}
\item state equation of the form \eqref{4.13},
\item dust matter fluid for $f_{1}=f_{3}+\theta\left(1+2\varphi\right)$,
\item radiation era for $f_{1}=f_{3}+\theta\left(1+4\varphi\right)$ and
\item dark energy epoch of the Universe for $f_{1}=f_{3}+\theta$.
\end{enumerate}
\end{cor}

\section{$P\left(GR_{4}\right)$ GRW spacetimes in $f(\mathcal{R},T^2)$ gravity}
The action for $f(\mathcal{R},T^2)$ gravity is described by
\begin{equation}\label{et1}
S=\int\left\{L_{m}+\dfrac{f\left(\mathcal{R},T^{2}\right)}{2\kappa}\right\}\sqrt{-g}d^{4}x
\end{equation}
where $L_{m}$ indicates the matter Lagrangian density depends on the metric $g_{ij}$ and $\kappa$ stands for coupling constant.\par
The action term yields the following field equations
\begin{equation}\label{e1}
f_{\mathcal{R}}\mathcal{R}_{jk}+\left\{g_{jk}\Box-\nabla_{j}\nabla_{k}\right\}f_{\mathcal{R}}+\Theta_{jk}f_{T^{2}}
=T_{jk}+\dfrac{1}{2}\,fg_{jk}
\end{equation}
in which $f_{\mathcal{R}}=\dfrac{\partial f}{\partial\mathcal{R}}$\,, $f_{T^{2}}=\dfrac{\partial f}{\partial T^{2}}$\,, $\Box\equiv\nabla_{l}\nabla^{l}$ and
\begin{equation}\label{e2}
\Theta_{jk}=\left\{Tg_{jk}-2T_{jk}\right\}L_{m}-4\dfrac{\partial^{2}L_{m}}{\partial g^{jk}\partial g^{li}}\,T^{li}+2T^{l}_{j}T_{jk}-TT_{jk}.
\end{equation}
When $f(\mathcal{R},T^2)= f(\mathcal{R})$, the field equation of this gravity turns into $f(\mathcal{R})$ theory, and when $f(\mathcal{R},T^2)= \mathcal{R}$, GR is retrieved.\par
Now, choose $L_{m}=p$ and the matter configuration as a perfect fluid. Then we have
\begin{equation}\label{e3}
\Theta_{jk}=-\left\{4p\sigma+3p^{2}+\sigma^{2}\right\}u_{j}u_{k}.
\end{equation}
Let us choose a model
\begin{equation}\label{e4}
f\left(\mathcal{R},T^{2}\right)=\mathcal{R}+T^{2}
\end{equation}
and also
\begin{equation}\label{e5}
T^{2}=T_{ij}T^{ij}=3p^{2}+\sigma^{2}.
\end{equation}
Multiplying equation \eqref{4.10} with $g^{jk}$, we reach
\begin{equation}\label{ee5}
\dfrac{f_{1}-f_{3}-\theta}{\varphi}=4\theta-\mathcal{R}.
\end{equation}
Utilizing \eqref{ee5} in \eqref{4.10}, we get
\begin{equation}\label{eee5}
\mathcal{R}_{jk}=\theta g_{jk}+\left(4\theta-\mathcal{R}\right)u_{j}u_{k}.
\end{equation}
Using the above equations in \eqref{e2}, we provide
\begin{equation}\label{e6}
\left(\dfrac{\mathcal{R}+3p^{2}+\sigma^{2}+2p-2\theta}{2}\right)g_{jk}+\left(3p^{2}
+4p\sigma+\sigma^{2}+p+\sigma-4\theta + \mathcal{R}\right)u_{j}u_{k}=0.
\end{equation}
Multiplying equation \eqref{e6} with $u^{j}$, we find
\begin{equation}\label{e7}
3p^{2}+8p\sigma+\sigma^{2}+2\sigma-6\theta+\mathcal{R}=0.
\end{equation}
Again, multiplying equation \eqref{e6} with $g^{jk}$ and using \eqref{ee5}, we have
\begin{equation}\label{e8}
3p^{2}-4p\sigma+\sigma^{2}-\sigma+3p+\mathcal{R}=0.
\end{equation}
Equations \eqref{e7} and \eqref{e8} give us
\begin{equation}\label{e9}
4p\sigma+\left(1-\kappa\right)\left(\sigma-p\right)=0,
\end{equation}
since $2\theta=\kappa(\sigma-p)$.\par
For %a stiff matter fluid, that is,
$\sigma=p$, the foregoing equation implies $p=\sigma=0$. Hence, from \eqref{1.2}, $T_{jk}=0.$ This infers that the spacetime is vacuum but recent experiment proves that a spacetime can not be vacuum since there always exist matter.\par
 If we consider $p=0$, then the above equation entails $\sigma=0$ and hence, $T_{jk}=0.$\par
For $\sigma=-p$, the foregoing equation implies $p=\sigma=$ constant.\par

Therefore, we obtain:
\begin{theo}
A $P\left(GR_{4}\right)$ GRW spacetime obeying $f\left(\mathcal{R},T^{2}\right)=\mathcal{R}+T^{2}$ gravity can not admit stiff matter fluid and dust matter fluid, but admit dark matter fluid with $p=\sigma=$ constant.
\end{theo}

\section{$P\left(GR_{4}\right)$ GRW spacetimes in Ricci tensor squared gravity}

Here, we investigate $f\left(\mathcal{R}^{\ast}\right)$-gravity model. For this gravity the action term,
\begin{equation}\label{e5.1}
S=\int\left\{L_{m}+\dfrac{\mathcal{R}+f\left(\mathcal{R}^{\ast}\right)}{2\kappa}\right\}d^{4}x\sqrt{-g}
\end{equation}
in which% $L_{m}$ stands for the matter Lagrangian density depends on the metric $g_{ij}$ and
 Ricci-tensor-squared gravity $\mathcal{R}^{\ast}$ is presented as
\begin{equation}\label{e5.2}
\mathcal{R}^{\ast}=\mathcal{R}_{jk}\mathcal{R}^{jk}.
\end{equation}
The variation of action term \eqref{e5.1} provides the modified EFE of $f\left(\mathcal{R}^{\ast}\right)$-gravity as \cite{LBM07}
\begin{equation}\label{e5.3}
\mathcal{R}_{jk}+2f_{\mathcal{R}^{\ast}}\mathcal{R}^{h}_{j}\mathcal{R}_{kh}-\dfrac{1}{2}\left\{\mathcal{R}
+f\left(\mathcal{R}^{\ast}\right)\right\}g_{jk}=\kappa T^{f}_{jk}
\end{equation}
in which $f_{\mathcal{R}^{\ast}}=\dfrac{\partial f}{\partial\mathcal{R}^{\ast}}$ and $T^{f}_{jk}$ is the EMT of the fluid.\par
In modified gravity the ECs are demonstrated as
\begin{align*}
\mathrm{Null\, Energy\, Condition} (\mathrm{NEC})&\quad\mathrm{if\;and\;only\;if}\quad\sigma+p\geq0,\\
\mathrm{Strong\, Energy\, Condition} (\mathrm{SEC})&\quad\mathrm{if\;and\;only\;if}\quad\sigma+p\geq0\quad\mathrm{and}\quad\sigma+3p\geq0,\\
\mathrm{Dominant\, Energy\, Condition} (\mathrm{DEC})&\quad\mathrm{if\;and\;only\;if}\quad\sigma\pm p\geq0\quad\mathrm{and}\quad\sigma\geq0,\\
\mathrm{Weak\, Energy\, Condition} (\mathrm{WEC})&\quad\mathrm{if\;and\;only\;if}\quad\sigma+p\geq0\quad\mathrm{and}\quad\sigma\geq0.
\end{align*}
Now, we choose PFS solutions in $f\left(\mathcal{R}^{\ast}\right)$-gravity equation assuming that the EMT takes the form \eqref{1.3}. Then, equations \eqref{1.3}, \eqref{4.1} and \eqref{e5.3} reflect that
\begin{align}\label{e5.4}
&\left\{\theta+2\theta^{2}f_{\mathcal{R}^{\ast}}-\dfrac{1}{2}\left(\mathcal{R}+f\left(\mathcal{R}^{\ast}\right)\right)
-\kappa p\right\}g_{jk}\nonumber\\
&+\left\{4\theta-\mathcal{R}-\left(16\theta^{2}-12\theta\mathcal{R}+2\mathcal{R}^{2}\right)f_{\mathcal{R}^{\ast}}
-\kappa p-\kappa\sigma\right\}u_{j}u_{k}=0.
\end{align}
Multiplying the above equation \eqref{e5.4} with $u^{j}$, we have
\begin{equation}\label{e5.5}
\kappa\sigma=3\theta-\mathcal{R}-2\left(3\theta-\mathcal{R}\right)^{2}f_{\mathcal{R}^{\ast}}
+\dfrac{1}{2}\left\{\mathcal{R}+f\left(\mathcal{R}^{\ast}\right)\right\}.
\end{equation}
Again, multiplying \eqref{e5.4} with $g^{jk}$ infers
\begin{equation}\label{e5.6}
3\kappa p-\kappa\sigma=\mathcal{R}+2\left\{12\theta^{2}-6\theta\mathcal{R}+\mathcal{R}^{2}\right\}f_{\mathcal{R}^{\ast}}
-2\left\{\mathcal{R}+f\left(\mathcal{R}^{\ast}\right)\right\}.
\end{equation}
From \eqref{e5.5} and \eqref{e5.6}, it follows that
\begin{equation}\label{e5.7}
\kappa p=\theta+2\theta^{2}f_{\mathcal{R}^{\ast}}-\dfrac{1}{2}\left\{\mathcal{R}+f\left(\mathcal{R}^{\ast}\right)\right\}.
\end{equation}
Therefore, we provide:
\begin{theo}
In a $P\left(GR_{4}\right)$ $\mathrm{GRW}$ spacetime solutions in $f\left(\mathcal{R}^{\ast}\right)$-gravity the energy density $\sigma$ and the isotropic pressure $p$ are described by \eqref{e5.5} and \eqref{e5.7}, respectively.
\end{theo}
\begin{rem}
Since $\sigma$ and $p$ are not constants, this result is consistent with the Cosmos as it exists right now.
\end{rem}
\noindent

From \eqref{eee5}, it follows that
\begin{equation}\label{5.8}
\mathcal{R}^{ij}=\theta g^{ij}+\left(4\theta-\mathcal{R}\right)u^{i}u^{j}.
\end{equation}
Equations \eqref{eee5} and \eqref{5.8} together imply
\begin{equation}\label{5.9}
\mathcal{R}_{ij}\mathcal{R}^{ij}=12\theta^{2}-6\theta\mathcal{R}+\mathcal{R}^{2}.
\end{equation}
Therefore the Ricci-tensor-squared gravity is
\begin{equation}\label{5.10}
\mathcal{R}^{\ast}=12\theta^{2}-6\theta\mathcal{R}+\mathcal{R}^{2}.
\end{equation}
In the following subsection, we verify the ECs for a $f\left(\mathcal{R}^{\ast}\right)$-gravity model.
\section*{{\bf A. $f\left(\mathcal{R}^{\ast}\right)=\ln(\mathcal{R}^{\ast})-\exp(-\mathcal{R}^{\ast})$}}
\noindent
With the help of \eqref{1.9}, \eqref{e5.5}, \eqref{e5.7} and \eqref{5.10}, $\sigma$ and $p$ are given by
\begin{align}\label{5.11}
\kappa\sigma=&3\theta-\dfrac{\mathcal{R}}{2}-\dfrac{2\left(3\theta-\mathcal{R}\right)^{2}}{12\theta^{2}-6\theta\mathcal{R}+\mathcal{R}^{2}}+\dfrac{1}{2}\ln(12\theta^{2}-6\theta\mathcal{R}+\mathcal{R}^{2})\nonumber\\
&-\left\{\dfrac{1}{2}+2\left(3\theta-\mathcal{R}\right)^{2}\right\}\exp(6\theta\mathcal{R}-12\theta^{2}-\mathcal{R}^{2}),
\end{align}
\begin{align}\label{5.12}
\kappa p=&\theta-\dfrac{\mathcal{R}}{2}+\dfrac{2\theta^{2}}{12\theta^{2}-6\theta\mathcal{R}+\mathcal{R}^{2}}-\dfrac{1}{2}\ln(12\theta^{2}-6\theta\mathcal{R}+\mathcal{R}^{2})\nonumber\\
&+\left(\dfrac{1}{2}+2\theta^{2}\right)\exp(6\theta\mathcal{R}-12\theta^{2}-\mathcal{R}^{2}).
\end{align}

The ECs for the above model are now examined. The ECs for this arrangement may now be discussed using equations \eqref{5.11} and \eqref{5.12}.\par
\begin{tabulary}{\linewidth}{CC}
	\includegraphics[height=0.24\textheight]{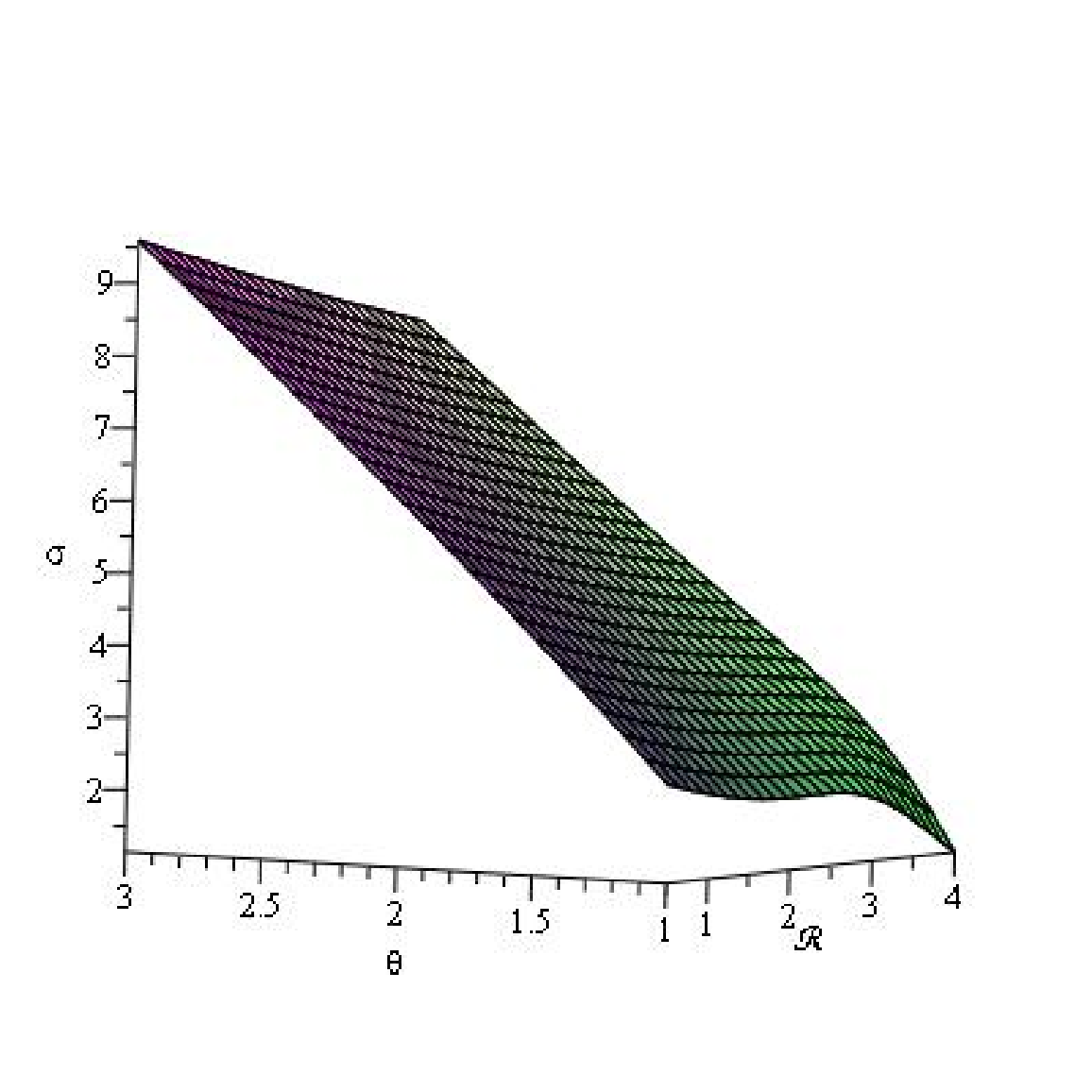}
	&
	\includegraphics[height=0.24\textheight]{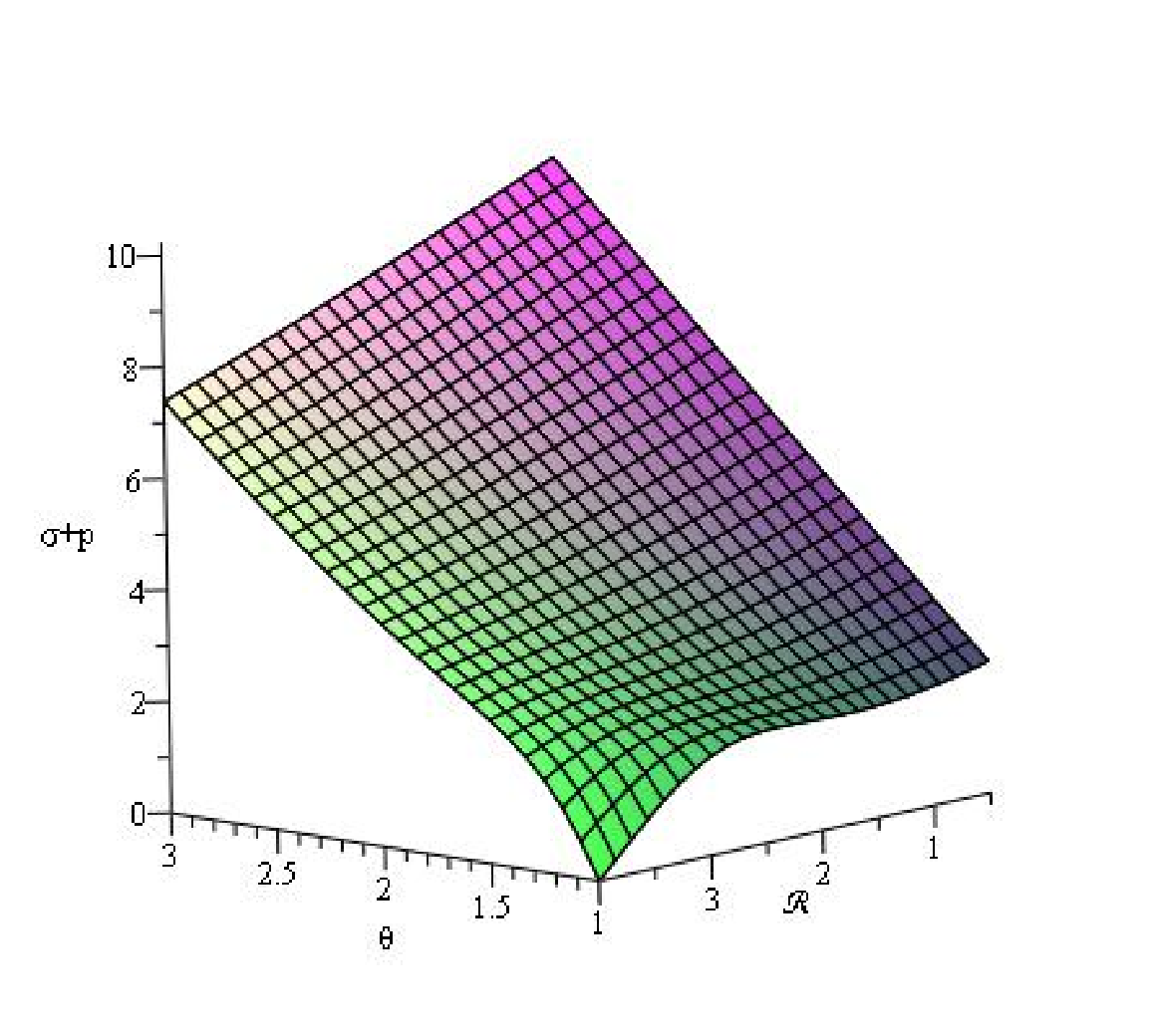}
	\\
	{\bf Fig. 1:} Development of $\sigma$ with reference to $\mathcal{R}$ and $\theta$ &{\bf Fig. 2:} Development of $p+\sigma$ with reference to $\mathcal{R}$ and $\theta$
	
\end{tabulary}
\begin{tabulary}{\linewidth}{CC}
	
	\includegraphics[height=0.25\textheight]{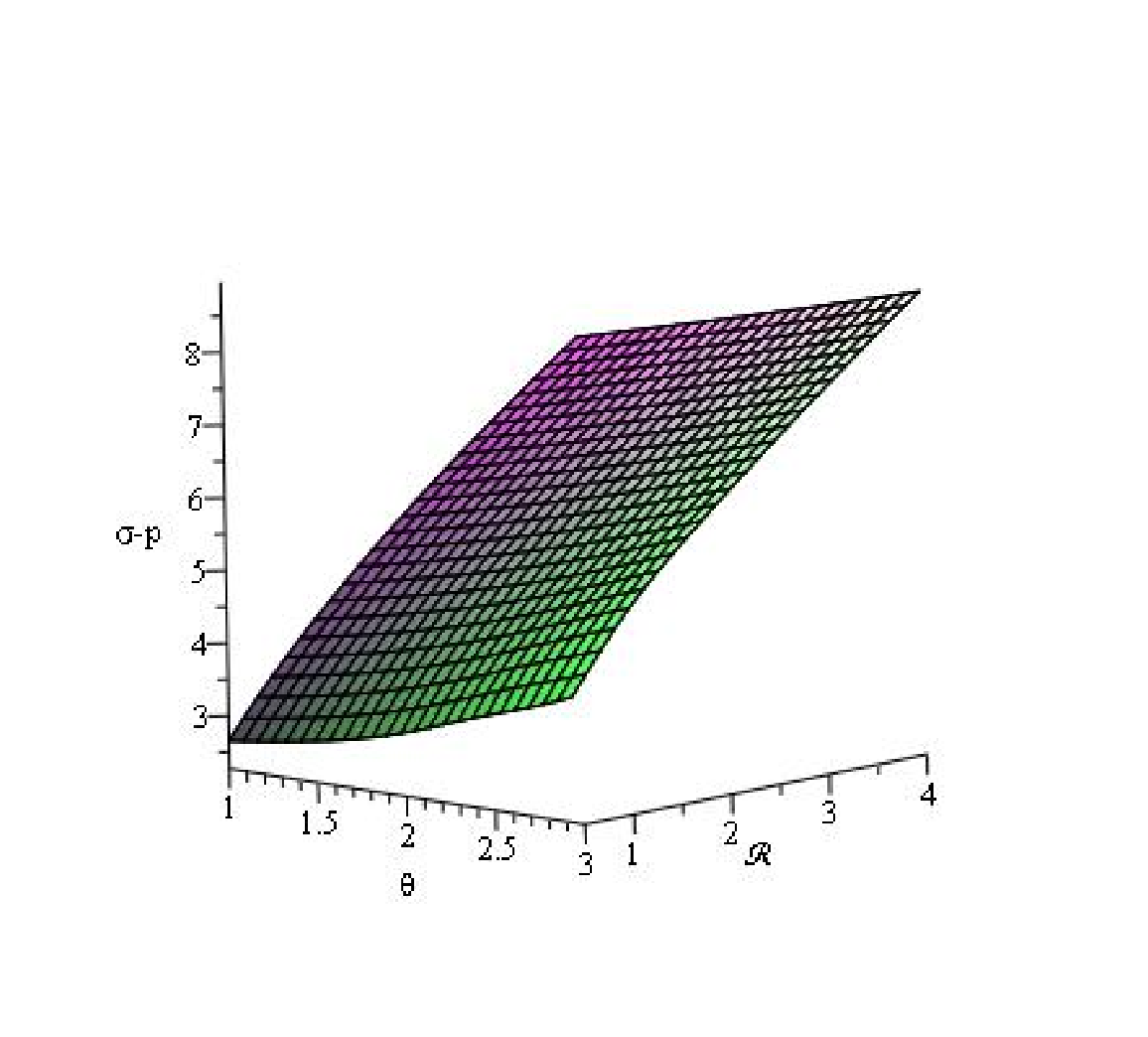}
    &
	\includegraphics[height=0.24\textheight]{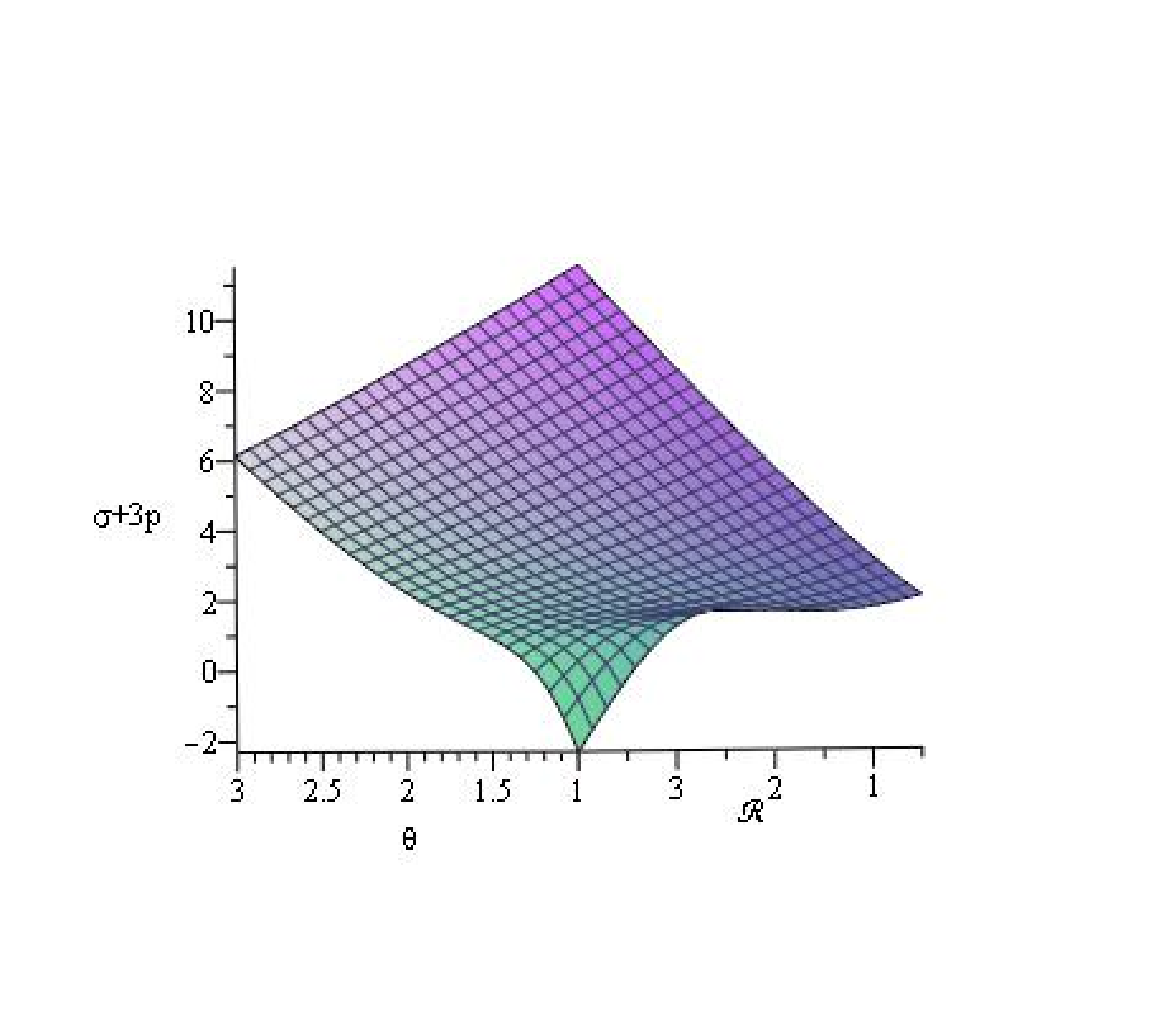}
    \\	
	
	{\bf Fig. 3:} Development of $\sigma-p$ with reference to $\mathcal{R}$ and $\theta$  &{\bf Fig. 4:} Development of $\sigma+3p$ with reference to $\mathcal{R}$ and $\theta$
	
\end{tabulary}

Figures $1$ and $2$ demonstrate that, for parameters $\theta, \mathcal{R}\in\left[1,3\right]$, the energy density and $p+\sigma$ can not be negative, and that, for larger values of $\theta$ and $\mathcal{R}$, they are high. Because NEC is a component of WEC, NEC and WEC are fulfilled. The $\sigma-\rho$ profile for $\theta, \mathcal{R}\in\left[1,3\right]$ is positive, as seen in Fig. $3$. It is evident from Figs. $1$, $2$, and $3$ that DEC is validated. Furthermore, we can observe that SEC is satisfied from Figs. $2$ and $4$, and this finding yields the late-time acceleration of the Cosmos\cite{LDMS21}. Moreover, each result aligns with the $\Lambda$CDM model \cite{AAAABBBBBBB20}.

\section{Discussion}
The physical motivation for researching various spacetime models in general relativity and cosmology is to gain additional insight into particular phases of the universe's evolution, which can be divided into the following three stages:\\
(i) The initial stage, (ii) The intermediate stage and (iii) The final stage.\par
The initial stage concerns about viscous fluid where as the intermediate stage concerns about non-viscous fluid and both are admitting heat flux. The final stage equipped with thermal equilibrium tells about perfect fluid stage.
In our current study, we select the final stage and it is shown that a $P\left(GR_{4}\right)$ $\mathrm{GRW}$ spacetime represents a PFS.\par
Here, with the geometric restriction of $P\left(GR_{4}\right)$ $\mathrm{GRW}$ spacetime, $f\left(\mathcal{R}^{\ast}\right)$ gravity model is investigated. Our findings have been assessed both graphically and analytically in this instance. To build our formulation and evaluate the model $f\left(\mathcal{R}^{\ast}\right)=\ln(\mathcal{R}^{\ast})-\exp(-\mathcal{R}^{\ast})$, we employed the analytical technique. Figs. $1$, $2$, $3$, and $4$ display the EC profiles for our model. For the parameters $\theta, \mathcal{R}\in\left[1,3\right]$, it has been found that the evolution of $\sigma$ is positive. Despite this, WEC, DEC, SEC, and NEC were all satisfied. These results, however, agree with the $\Lambda$CDM model.\par

\section{Declarations}
\subsection{Funding }
NA.
\subsection{Code availability}
NA.
\subsection{Availability of data}
NA.
\subsection{Conflicts of interest}
The authors have no conflicts to disclose.

\end{document}